\documentclass[aps,amsfonts,prd,nofootinbib,showpacs,showkeys]{revtex4}

\usepackage{epsfig}

\newcommand{\be}{\begin{eqnarray}}
\newcommand{\ee}{\end{eqnarray}}

\begin{document}

\title{Exotic Baryons and Monopole Excitations
in a Chiral Soliton Model}

\author{H.~Weigel}

\affiliation{Fachbereich Physik, Siegen University\\
Walter--Flex--Stra{\ss}e 3, D--57068 Siegen, Germany}

\begin{abstract}
\baselineskip16pt
We compute the spectra of exotic pentaquarks and monopole excitations 
of the low--lying $\frac{1}{2}^+$ and $\frac{3}{2}^+$ baryons in
a chiral soliton model. Once the low--lying baryon properties
are fit, the other states are predicted without any more 
adjustable parameters. This approach naturally leads to a 
scenario in which the mass spectrum of the next to lowest--lying 
$J^\pi=\frac{1}{2}^+$ states is fairly well approximated by the 
ideal mixing pattern of the 
$\mathbf{8}\oplus\overline{\mathbf{10}}$
representation of flavor $SU(3)$. We compare our results to predictions
obtained in other pictures for pentaquarks and speculate about
the spin--parity assignment for~$\Xi(1690)$ and~$\Xi(1950)$.
\end{abstract}

\maketitle

\baselineskip16pt
\section{Introduction}

Although chiral soliton model predictions for the mass of the lightest 
exotic pentaquark, the $\Theta^+$ with zero isospin and unit strangeness, 
have been around for about two decades~\cite{early}, the study of such 
pentaquarks as potential baryon resonances became popular only very 
recently when experiments~\cite{Thetaexp,NA49} indicated their existence. 
These experiments were stimulated by a chiral soliton model 
estimate that suggested~\cite{Di97} that such exotic baryons might 
have a width\footnote{In chiral soliton models the direct extraction of
the interaction Hamiltonian for hadronic decays of resonances still 
is an unresolved issue. Estimates are obtained from axial current 
matrix elements~\cite{Di97,We98,Pr03,Ja04}. In view of what is 
known about the related $\Delta\to\pi N$ transition matrix 
element~\cite{deltadecay}, such estimates may be questioned. It is also
worth noting that the computation of this particular decay width
has a long and notable history. The interested reader may trace it 
from ref.~\cite{Ja04}.} that is much smaller than those 
typical for hadronic decays of baryon resonances~\cite{Di97,Ja04}.
Such narrow resonances could have escaped detection in earlier 
analyses. Then very quickly the experimental observations initiated 
exhaustive studies on the properties of pentaquark baryons. Comprehensive 
lists of such studies are, for example, collected in refs.~\cite{Je03,El04}.

\begin{figure}[t]
\parbox[t]{1cm}{~}
\parbox[t]{6cm}{
\epsfig{file=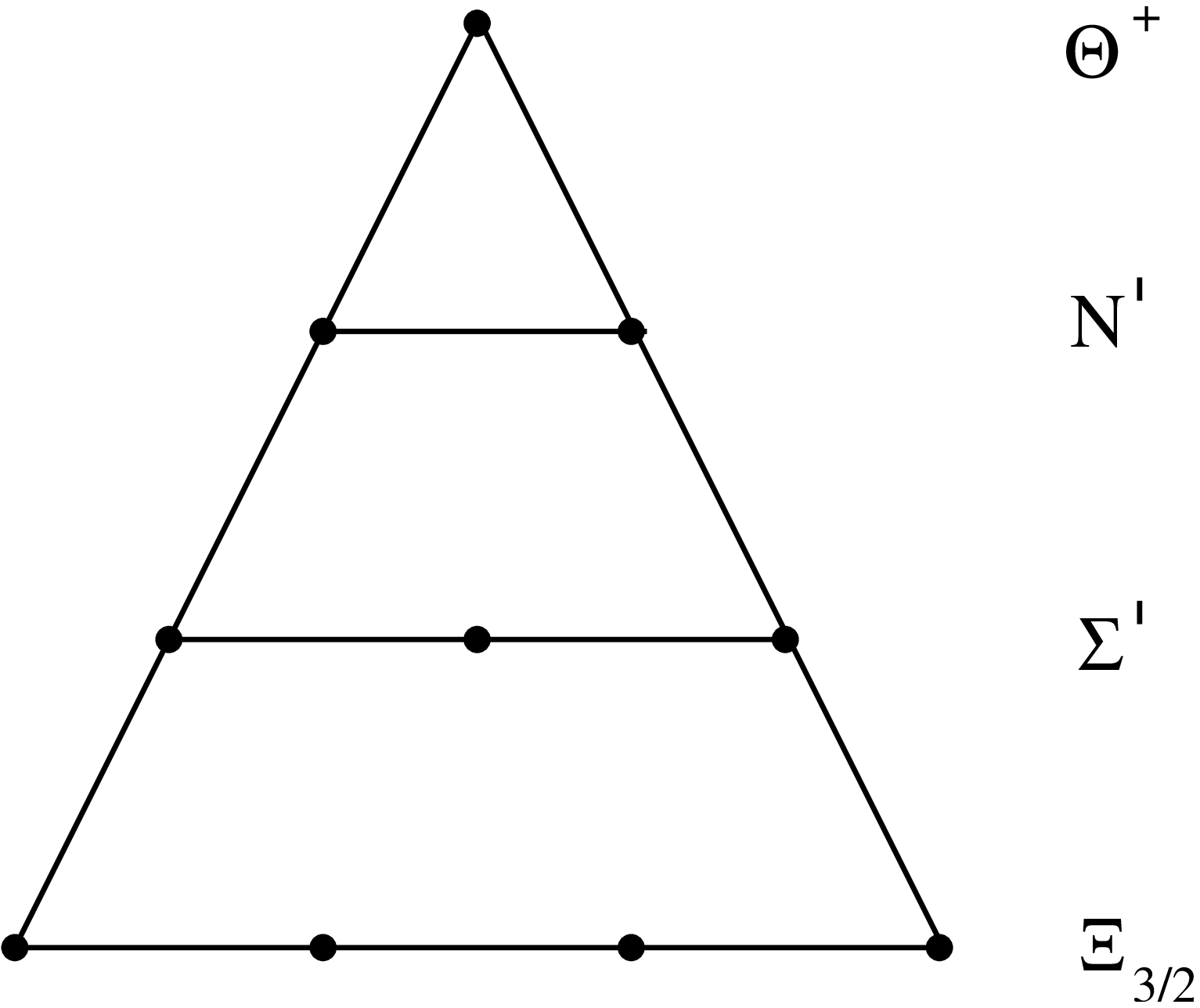,height=3.8cm,width=5cm}}
\parbox[t]{2cm}{~}
\parbox[t]{7.0cm}{
\epsfig{file=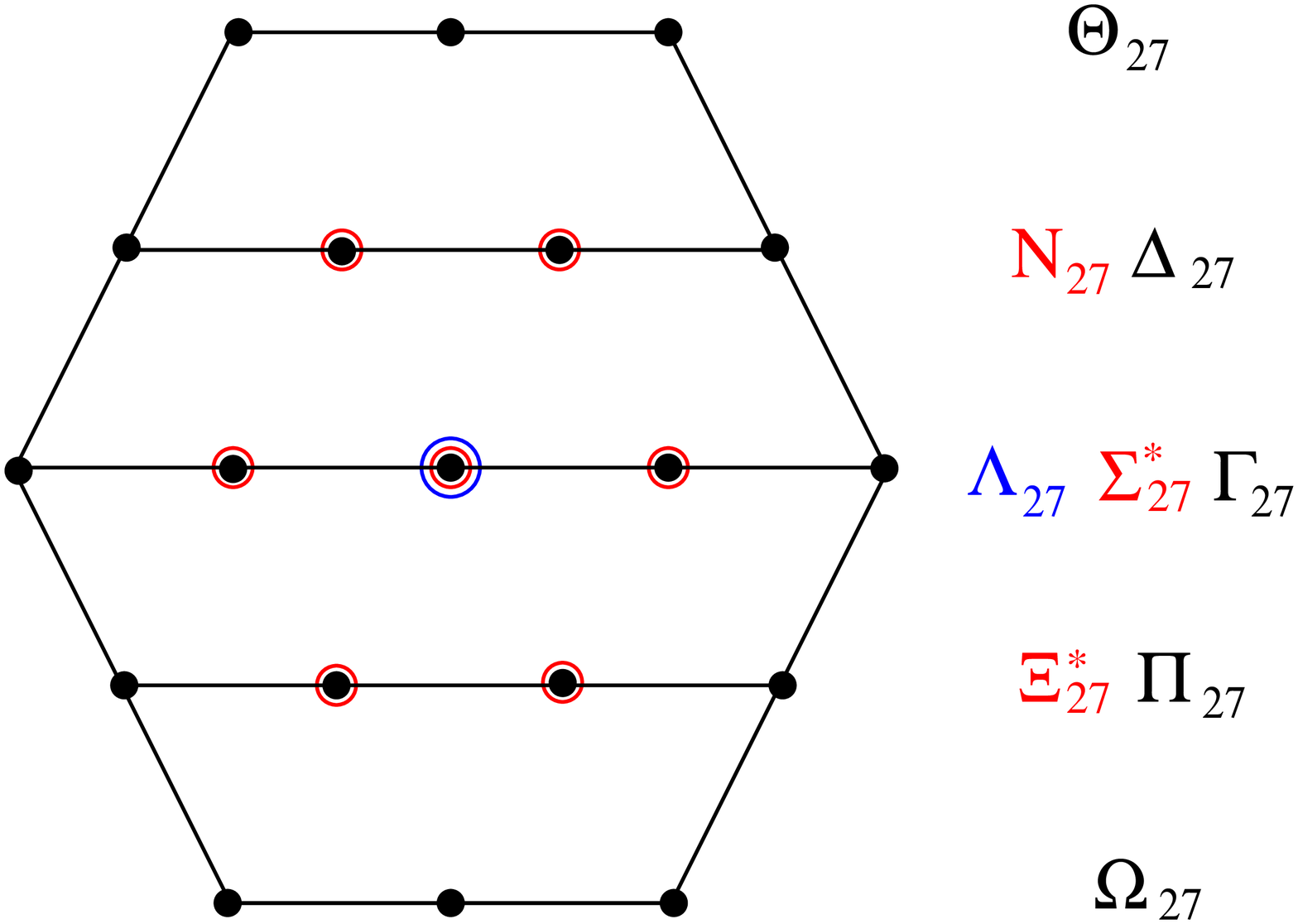,height=5cm,width=6.5cm}}
\parbox[t]{1cm}{~}
\caption{\label{fig_1}A sketch of the $SU_F(3)$ representations
$\overline{\mathbf{10}}$ and $\mathbf{27}$ with their 
exotic baryon members: $\Theta^+,\Xi_{3/2},\Theta_{27},
\Lambda_{27},\Gamma_{27},\Pi_{27}$ and~$\Omega_{27}$.
As usual various isospin projections are plotted horizontally 
while states of different hypercharge are spread vertically.}
\end{figure}

Chiral soliton models are a common platform for such studies because 
higher dimensional $SU_F(3)$ irreducible representations that 
contain exotic pentaquarks emerge almost naturally.
In these models states with baryon 
quantum numbers are generated from the soliton by canonically 
quantizing the collective coordinates that parameterize the large 
amplitude fluctuations associated with (would--be) zero modes. 
When extending the model to $SU_F(3)$ the lowest states are 
members of the flavor octet and decuplet representations 
for $J^\pi=\frac{1}{2}^+$ and $J^\pi=\frac{3}{2}^+$, 
respectively. Upon inclusion of flavor symmetry breaking the 
physical states acquire admixtures from higher dimensional 
representations. For the $J^\pi=\frac{1}{2}^+$ baryons those 
admixtures originate dominantly from the antidecuplet, 
$\overline{\mathbf{10}}$, and the 
$\mathbf{27}$--plet~\cite{Gu84,We96,Ya88,Pa89}. The particle 
content of these representations is depicted in figure~\ref{fig_1}.
In addition to states with quantum numbers of octet baryons these
representations obviously also contain states with quantum numbers 
that \emph{cannot} be built as three quark composites. These 
states contain (at least) one additional quark--antiquark pair. 
Hence the notion of exotic pentaquarks.
So far, the $\Theta^+$ and $\Xi_{3/2}$ with masses of 
$1537\pm10{\rm MeV}$~\cite{Thetaexp} and $1862\pm2{\rm MeV}$~\cite{NA49}
have been observed, even though the single observation of $\Xi_{3/2}$ is 
not undisputed~\cite{Fi04} and awaits confirmation. In soliton models 
these two resonances are identified as members of the antidecuplet 
$SU_F(3)$ representation. Therefore they should carry the quantum 
numbers $I(J^\pi)=0(\frac{1}{2}^+)$ for $\Theta^+$ and
$\frac{3}{2}(\frac{1}{2}^+)$ for $\Xi_{3/2}$. 
We stress that these quantum numbers are not yet confirmed
by experiment. The channels for which the resonance analyses were 
performed are $nK^+$ and $pK^0_S$ for $\Theta^+$~\cite{Thetaexp} and 
$\Xi^-\pi^-$ for $\Xi_{3/2}$~\cite{NA49}. Hence the strangeness quantum 
numbers of these exotics are without ambiguities: $S(\Theta^+)=1$ and
$S(\Xi_{3/2})=-2$ and the isospin of $\Xi_{3/2}$ should at least 
be $\frac{3}{2}$.

It is now suggestive to wonder about the nature of states in those 
higher dimensional representation that do carry quantum numbers
of three quark composites. In the antidecuplet these are $N^\prime$
and $\Sigma^\prime$, {\it cf.} figure~\ref{fig_1}. Similarly,
$\Delta_{27}$, $\Sigma_{27}$ and $\Xi_{27}$ have the same quantum 
numbers as the ordinary $\Delta$, $\Sigma^*$ and $\Xi^*$ from the
decuplet, respectively. In ref.~\cite{Di97}, for example,
the $N^\prime \in \overline{\mathbf{10}}$ was identified
as the $N(1710)$ resonance to adjust the overall mass scale 
of the antidecuplet. Though that approach correctly predicted 
the mass of $\Theta^+$, the prediction for $\Xi_{3/2}$ turned 
out to be incompatible with the observation. It was then quickly 
realized that the $N^\prime \in \overline{\mathbf{10}}$ and
$\Sigma^\prime \in \overline{\mathbf{10}}$ could easily 
mix with nucleon and $\Sigma$ states that are members 
of an additional octet (besides the ground state octet). 
This picture leads to the description of pentaquarks as members
of the direct sum $\mathbf{8}\oplus\overline{\mathbf{10}}$ 
and could be motivated both in quark--diquark~\cite{Ja03}
and chiral soliton~\cite{Di03} approaches. However, the latter
description did not provide any dynamical origin for the 
additional octet. Of course, it is very
natural to view this additional octet as a monopole (or radial)
excitation of the ground state octet. The fact that
such monopole excitations would mix with the corresponding
states in the antidecuplet was recognized already some time 
ago~\cite{Sch91,Sch91a} and a dynamical model was 
developed not only to investigate such mixing effects 
but also to describe static properties of the low--lying 
$J^\pi=\frac{1}{2}^+$ and $J^\pi=\frac{3}{2}^+$ baryons.
Technically that was accomplished by not only quantizing
the collective coordinates that parameterize the flavor
(and spatial) orientation of the soliton but also its
spatial extension. This approach was further justified by the 
observation that the proper description of baryon magnetic moments 
requires a substantial feedback of flavor symmetry breaking on 
the soliton size~\cite{Schw91}. 

The philosophy of chiral soliton models is to first construct 
a chiral Lagrangian for meson fields and determine as many 
model parameters as possible from meson properties. Baryons 
then emerge as solitons of this meson theory making the approach 
very predictive in the baryon sector. In the context of pentaquark 
studies, however, a commonly adopted (model independent) strategy is 
to write down all operators in the space of collective coordinates 
up to a given order in $1/N_C$ and/or flavor symmetry breaking 
(measured by the strange quark mass) and determine the unknown 
coefficients from baryon properties~\cite{Di97,El04,Wa03,Bo03,Wu03}. 
Once the additional collective coordinate for the monopole excitation 
is included, the number of symmetry--allowed operators in the Hamiltonian 
is no longer limited to just a few. Therefore such model independent 
approaches turn out not to be very predictive because not sufficient 
information is available to determine all coefficients. Rather a fully 
dynamical calculation must be performed in a given model. This has the 
additional advantage that no assumption is used as input about the nature 
of states in the higher dimensional $SU_F(3)$ representations. In 
particular, no such input is necessary to fit the mass scale for 
states in the $\overline{\mathbf{10}}$ and $\mathbf{27}$ 
representations.

In the studies of refs.~\cite{Sch91,Sch91a} the spectra for radial 
excitations of the low--lying~$\frac{1}{2}^+$ and $\frac{3}{2}^+$ 
baryons were computed in the framework of an $SU_F(3)$ chiral 
soliton model. In addition the version of the model 
discussed in ref.~\cite{Sch91a} was shown to describe the static 
properties of those baryons fairly well. Later the mass of 
the recently discovered $\Theta^+$ pentaquark was predicted with 
reasonable accuracy in the same model~\cite{We98,We00}. Here we will 
therefore employ exactly that model without any further modifications to 
predict masses of the exotic $\Xi_{3/2}$ and of additional exotic 
baryons that originate from the $SU_F(3)$ representations 
$\overline{\mathbf{10}}$ and~$\mathbf{27}$, respectively. The latter 
are exotic pentaquarks with $J^\pi=\frac{3}{2}^+$ and may be considered 
as partners of $\Theta^+$ and $\Xi_{3/2}$ in the same way as
the $\Delta$ is the partner of the nucleon. We will also reanalyze 
the $J^\pi=\frac{1}{2}^+$ excited baryons in view of the recently
developed $\mathbf{8}\oplus\overline{\mathbf{10}}$ scenario. 

Essentially the model has only a single free parameter that has 
earlier~\cite{Sch91,Sch91a} been fixed to provide an acceptably 
nice picture of baryon properties. In this way, it is also possible 
to discuss the relevance of mixing between rotational and vibrational 
modes when attempting to predict the spectrum of pentaquarks. This is 
crucial for pentaquarks because the mass difference to the nucleon of 
these modes are of the same order in the $1/N_C$ expansion~\cite{Kl89}. 
Hence mixing effects are potentially large even they are mediated by 
the (small) flavor symmetry breaking.

The paper is organized as follows. In section II we will review the 
specific soliton model. As will be described in section III the 
quantization of both the rotational and the monopole degrees of 
freedom leads to the baryon states that mix with the (exotic) 
pentaquarks in the antidecuplet and the $\mathbf{27}$--plet.
The numerical results on the spectrum will be presented 
in section IV. We compare our results to those obtained in
other descriptions for the pentaquark spectrum that are also
formulated in the framework of mixing between excited octet 
and antidecuplet baryons in section V. We summarize 
in section VI. A final word on notation. We will call exotics 
or exotic pentaquarks all baryons that must contain at least one 
antiquark to saturate the quantum numbers. On the other hand we
call pentaquarks all baryonic members of $SU_F(3)$ representations
whose Young tableaux need one antiquark even though the quantum numbers
of some of these baryons can be built from three quark states; in
this case they are non--exotic.

\section{The Soliton Model}

In this section we will describe the soliton model that we will 
use later to analyze monopole and rotational degrees of freedom.
For the present investigation we will employ the Skyrme model 
extended by a scalar field as motivated by the trace anomaly 
of QCD. There are reasons to believe that extensions with 
vector mesons or chiral quarks are more realistic~\cite{We96}.
However, and as discussed in ref.~\cite{We98} technical subtleties 
may occur when quantizing the radial degrees of freedom in more
realistic soliton models like those with chiral quarks and/or vector
mesons. We therefore consider the scalar meson extension of
the Skyrmion as adequate.

To be specific we will follow the treatment of ref.~\cite{Sch91a} where 
the soliton model contains a scalar meson, $\sigma$ and eight pseudoscalar 
mesons~$\phi^a$, $a=1,\ldots,8$. The scalar meson field parameterizes
$H=\langle H\rangle{\rm exp}(4\sigma)
=\frac{\beta(g)}{g} G_{\mu\nu}^a G^{a\mu\nu}$ which
is introduced as an effective order parameter for the gluon field
to imitate the dilation anomaly equation of QCD~\cite{Go86},
$-\partial_\mu D^\mu= H + \sum_i m_i{\bar \Psi}_i\Psi_i$.
The vacuum expectation value $\langle H \rangle\sim
(0.30-0.35 {\rm GeV})^4$ can be extracted from sum rule estimates for the
gluon condensate~\cite{Sh79}. Eventually the fluctuating field $\sigma$
may be identified as a scalar glueball. The effective mesonic action reads
\be
\Gamma=\int d^4x \left({\cal L}_0+{\cal L}_{\rm SB}\right)
+\Gamma_{\rm WZ} \ .
\label{act1}
\ee
The flavor symmetric part involves both the chiral field\footnote{Here 
the normalization coefficients $f_a$ refer to the pseudoscalar decay 
constants.} $U={\rm exp} (i\lambda^a\phi^a/f_a)$ 
as well as the scalar fluctuation $\sigma$
\be
{\cal L}_0&=&-\frac{f_\pi^2}{4}{\rm e}^{2\sigma}
{\rm tr}\left(\alpha_\mu\alpha^\mu\right)
+\frac{1}{32e^2}{\rm tr}
\left(\left[\alpha_\mu,\alpha_\nu\right]^2\right)
+\frac{1}{2}\Gamma_0^2\ {\rm e}^{2\sigma} \ 
\partial_\mu\sigma\partial^\mu\sigma
+{\rm e}^{4\sigma}\left\{\frac{1}{4}
\left[\langle H \rangle 
-6\left(2\delta^\prime+\delta^{\prime\prime}\right)\right]
- \sigma\langle H \rangle\right\}\,,
\label{act2}
\ee
with $\alpha_\mu=\partial_\mu U U^\dagger$. Assuming the canonical 
dimensions $d(U)=0$ and $d(H)=4$ it is straightforward to verify 
that the Lagrangian~(\ref{act2}) yields the dilation anomaly equation for 
$m_i=0$. The terms which lift the degeneracy between the pseudoscalar
mesons of different strangeness are comprised in the flavor 
symmetry breaking part of the Lagrangian,
\be
{\cal L}_{\rm SB}={\rm tr}\Big\{
\left(\beta^\prime{\hat T}+\beta^{\prime\prime}{\hat S}\right)
{\rm e}^{2\sigma}\partial_\mu U \partial^\mu U^\dagger U
+\left(\delta^\prime{\hat T}+\delta^{\prime\prime}{\hat S}\right)
{\rm e}^{3\sigma}U+{\rm h. c.}\Big\}\ ,
\label{act3}
\ee
where the flavor projectors ${\hat T}={\rm diag}(1,1,0)$ and 
${\hat S}={\rm diag}(0,0,1)$ have been introduced. The coupling of 
the scalar field in ${\cal L}_{\rm SB}$ is such as to reproduce the 
explicit breaking in the dilation anomaly equation~\cite{Ad87}. 
The various parameters in eqs.~(\ref{act2}) and (\ref{act3}) are 
determined from the well--known masses and decay constants of the 
pseudoscalar mesons:
\be
\beta^\prime\approx 26.4 {\rm MeV}^2,\
\beta^{\prime\prime}\approx 985 {\rm MeV}^2,\
\delta^\prime\approx 4.15\times10^{-5} {\rm GeV}^2,\
\delta^{\prime\prime}\approx 1.55\times10^{-3}{\rm GeV}^4\ .
\label{para1}
\ee
Then the only free parameters of the model are the Skyrme 
constant $e$ and the glueball mass,
\be
m_\sigma^2=\frac{4\langle H \rangle +
6(2\delta^\prime+\delta^{\prime\prime})}{\Gamma_0^2} \ .
\label{msig}
\ee
As in ref.~\cite{Sch91a} we will always take 
$m_\sigma\approx1.25{\rm GeV}$. This corresponds to 
$\langle H \rangle=0.01{\rm GeV}^4$ and $\Gamma_0=180{\rm MeV}$.
Finally the scale invariant Wess--Zumino term is most conveniently 
presented by introducing the one--form $dUU^\dagger=\alpha_\mu dx^\mu$,
\be
\Gamma_{\rm WZ}=\frac{iN_c}{240\pi^2}\int 
{\rm tr}\left[(dUU^\dagger)^5\right] \, .
\label{wzterm}
\ee
The above described model possesses a static soliton solution 
of hedgehog structure
\be
U_0(\mbox{\boldmath $r$})={\rm exp}
[i \mbox{\boldmath $\tau$}\cdot \hat{\mbox{\boldmath $r$}}F(r)]
\qquad {\rm and} \qquad
\sigma(\mbox{\boldmath $r$})=\sigma_0(r) 
\label{soliton}
\ee
which is characterized by the two radial functions $F(r)$ and 
$\sigma_0(r)$~\cite{Go86,Ja87}. This configuration has non--trivial 
topology and the corresponding winding number is identified with 
the baryon charge. The profile functions $F(r)$ and $\sigma_0(r)$ are 
determined from the Euler--Lagrange equations to the action~(\ref{act1})
with boundary conditions suitable for baryon number one.

\section{Quantization of the soliton}

Besides baryon number the configuration~(\ref{soliton}) does not 
carry quantum numbers of physical baryon states such as spin or 
isospin. These are generated by canonical quantization of the 
collective coordinates which are introduced as time dependent
parameters to describe large amplitude fluctuations around the 
soliton. Apparently these are the rotations in coordinate and 
flavor spaces. The hedgehog configuration~(\ref{soliton}) is embedded 
in the isospin space. Hence the coordinate rotations may be
expressed as rotations in the flavor subspace of isospin and it 
sufficies to consider flavor rotations.  These rotational modes 
are parameterized by $A(t)\in SU_F(3)$ and their
quantization will thus lead to states that carry good spin and 
flavor quantum numbers. In addition the energy surface associated 
with scale or breathing transformations of the soliton is known to 
be shallow, at least in a large vicinity of the stationary 
point~\cite{Ha84,HH84} allowing for large amplitude fluctuations. 
This is even more the case when the dilatation symmetry is respected. 
For this reason it is suggestive to also introduce a collective 
coordinate, $\mu(t)$ for the spatial extension of the soliton. The 
quantization of this mode will then describe the monopole excitations. 
We therefore consider the following time dependent meson configuration 
that emerges from the soliton~(\ref{soliton}),
\be
U(\mbox{\boldmath $r$},t) = 
A(t) U_0\left(\mu(t)\mbox{\boldmath $r$}\right) A^\dagger(t)
\quad {\rm and}\quad
\sigma(\mbox{\boldmath $r$},t)=\sigma_0\left(\mu(t)r\right) \ .
\label{colcoor}
\ee
We then substitute this parameterization into the action (\ref{act1})
and obtain the Lagrangian for the collective rotations $A(t)$ and
the monopole vibration $x(t)=[\mu(t)]^{-3/2}$
\be
L(x,\dot x,A,\dot A)\hspace{-0.1cm}&=&\hspace{-0.1cm}{4\over9}
\left(a_1+a_2x^{-{4\over3}}\right){\dot x^2}-
\left(b_1x^{2\over3}+b_2x^{-{2\over3}}+b_3x^2\right)
+{1\over2}\left(\alpha_1x^2+\alpha_2x^{2\over3}\right)
\sum^3_{a=1}\Omega^2_a\hspace{1cm}
\nonumber \\* && 
+{1\over2}\left(\beta_1x^2+\beta_2x^{2\over3}\right)
\sum^7_{a=4}\Omega^2_a +{\sqrt3\over2}\Omega_8
-\left(s_1x^2+s_2x^{2\over3}+{4\over9}s_3{\dot x^2}\right)
\left(1-D_{88}\right) \ .
\label{lagbreath}
\ee
Here we have introduced the angular velocities 
$A^\dagger \dot A = (i/2)\sum_{a=1}^8 \lambda_a\Omega_a$ as well as 
the adjoint representation 
$D_{ab}=(1/2){\rm tr}(\lambda_a A \lambda_b A^\dagger)$. A term linear 
in $\dot x$, which would originate from flavor symmetry breaking terms, 
has been omitted because the matrix elements of the associated $SU(3)$ 
operators vanish when properly accounting for Hermiticity in the process 
of quantization \cite{Pa91}. The expressions for the constants 
$a_1,\ldots,s_3$ are functionals of the profile functions $F(r)$ and 
$\sigma_0(r)$. In refs.~\cite{Sch91,Sch91a} their analytic 
expressions and numerical values are given as well as a discussion
on the approximate treatment of ontributions from ${\cal L}_{\rm SB}$ 
that are quadratic in the angular velocities. As far as the rotational 
collective coordinates are concerned, the Lagrangian~(\ref{lagbreath}) 
contains only a single flavor symmetry breaking operator, $1-D_{88}$. 
Many other soliton model 
analyses consider up to three such operators~\cite{Di97,Di03,Wa03,Bo03,Wu03}. 
Rather than determining the coefficients of these operators in an
actual soliton model calculation they are frequently adjusted to baryon
properties. The additional operators are subleading in $1/N_C$ and 
do not emerge in models that have time derivatives appearing only 
with even powers in the flavor symmetry breaking terms of the 
Lagrangian, such as in eq.~(\ref{act3}). On the other hand vector 
meson~($V_\mu \partial^\mu U U^\dagger$) or chiral 
quark~($\bar{\Psi}\gamma_\mu\partial^\mu \Psi$) model extensions may 
lead to such subleading operators. Explicit calculations for the 
coefficients of these operators in appropriate models indicate~\cite{We96}, 
however, that the inclusion of such subleading operators affects 
the resulting mass spectrum only slightly.

We are now in the position to canonically quantize the collective
coordinates by constructing the Hamiltonian from the 
Lagrangian~(\ref{lagbreath}). For convenience we first define
\be
m&=&m(x)=\frac{8}{9}(a_1+a_2x^{-{4\over3}})\ , \quad
b=b(x)=b_1x^{2\over3}+b_2x^{-{2\over3}}+b_3x^2\,,
\nonumber \\
\alpha&=&\alpha(x)=\alpha_1x^2+\alpha_2 x^{\frac{2}{3}}\ , \quad
\beta=\beta(x)=\beta_1x^2+\beta_2 x^{\frac{2}{3}}\,,
\nonumber
\ee
and
\be
s&=&s(x)=s_1x^2+s_2x^{\frac{2}{3}} \ .
\label{defbreath}
\ee
The resulting Hamiltonian for the collective coordinates
consists of a flavor symmetric and a flavor symmetry breaking
piece. The flavor symmetric part of the collective Hamiltonian
\be
H_{\rm sym}=-\frac{1}{2\sqrt{m\alpha^3\beta^4}}\frac{\partial}{\partial x}
\sqrt{\frac{\alpha^3\beta^4}{m}}\frac{\partial}{\partial x}
+b+\left(\frac{1}{2\alpha}-\frac{1}{2\beta}\right)\mathbf{J}^2
+\frac{1}{2\beta}C_2-\frac{3}{8\beta}+s
\label{freebreath}
\ee
contains the collective rotations $A$ only through their canonical 
momenta, $R_a=-\frac{\partial L}{\partial \Omega_a}$. These are the 
spin operator, $\mathbf{J}^2=\sum_{i=1}^3R_i^2$, and the quadratic 
Casimir operator, $C_2=\sum_{a=1}^8R_a^2$, of $SU_F(3)$. These operators 
are diagonal for a definite $SU(3)$ representation, $\mu$. Due to the 
hedgehog structure of the static configuration $U_0$ and $\sigma_0$, 
the allowed representations must contain at least one state with 
identical spin and isospin. In addition, the constraint $R_8=-\sqrt{3}/2$,
that originates from the Wess--Zumino term~(\ref{wzterm}), requires
this state to possess unit hypercharge and only permits states 
with half--integer spin~\cite{early,Gu84,We96}. Then the allowed 
$SU(3)$ representations are 
$\mu=\mathbf{8}$, $\mu=\overline{\mathbf{10}}$, $\mu=\mathbf{27}$
$\mu=\overline{\mathbf{35}}$, etc. for spin $J=\frac{1}{2}$ and
$\mu=\mathbf{10}$, $\mu=\mathbf{27}$, $\mu=\mathbf{35}$,
$\mu=\overline{\mathbf{35}}$, etc. spin $J=\frac{3}{2}$. 
When we substitute the quantum numbers $J(J+1)$ and 
$C_2(\mu)$ [$C_2(\mathbf{8})=3$, 
$C_2(\mathbf{10})=C_2(\overline{\mathbf{10}})=6$, 
$C_2(\mathbf{27})=8$, etc.], $H_{\rm sym}$ is as simple as an 
ordinary second order differential operator in the monopole mode $x$.
This Schr\"odinger equation can straightforwardly be solved
numerically. For definiteness we denote the eigenvalues by 
${\cal E}_{\mu,n_\mu}$ and the corresponding eigenstates 
by $|\mu,n_\mu\rangle$, where $n_\mu$ labels the monopole (or radial) 
excitations for a prescribed $SU(3)$ representation, $\mu$. 
According to the above discussion the eigenstates factorize, {\it i.e.}
$|\mu,n_\mu\rangle=|\mu\rangle|n_\mu\rangle$. In this language the
nucleon corresponds to $|{\bf 8},0\rangle$ while the first radially 
excited state, which is commonly identified with the Roper~(1440) 
resonance, would be $|{\bf 8},1\rangle$. We are interested in the role 
of states like $|{\overline {\bf 10}},n_{\overline {\bf 10}}\rangle$ 
since in particular this tower contains the state with the quantum 
numbers of the exotic pentaquarks $\Theta^+$ and $\Xi_{3/2}$. Their 
partners with spin $J=3/2$ are contained in 
$|\mathbf{27},n_{\mathbf{27}}\rangle$.

In the second step we include the symmetry breaking part of the
Hamiltonian obtained canonically from the Lagrangian~(\ref{lagbreath}).
We utilize the above constructed states $|\mu,n_\mu\rangle$ as
a basis to diagonalize the complete Hamiltonian matrix
\be
H_{\mu,n_\mu;\mu^\prime,n^\prime_{\mu^\prime}}=
{\cal E}_{\mu,n_\mu}\delta_{\mu,\mu^\prime}
\delta_{n_\mu,n^\prime_{\mu^\prime}}
-\langle\mu|D_{88}|\mu^\prime\rangle
\langle n_\mu|s(x)|n^\prime_{\mu^\prime}\rangle \ .
\label{hammatr}
\ee
The flavor part of these matrix elements is computed using $SU(3)$ 
Clebsch--Gordan coefficients while the radial part is calculated numerically
using the appropriate eigenstates of (\ref{freebreath}). Of course, this 
can be done for each isospin multiplet separately, {\it i.e.} flavor 
quantum numbers are not mixed. We stress that we \emph{exactly} 
diagonalize the complete Hamiltonian, $H_{\rm sym}-s(x)D_{88}$ rather 
than approximating the eigenvalues and eigenstates in form of a 
perturbation in flavor symmetry breaking.
The physical baryon states $|B,m\rangle$ are 
finally expressed as linear combinations of the eigenstates of
the symmetric part
\be
|B,m\rangle=\sum_{\mu,n_\mu}C_{\mu,n_\mu}^{(B,m)}
|\mu,n_\mu\rangle \ .
\label{bsbreath}
\ee
The corresponding eigen--energies are denoted by $E_{B,m}$.
The nucleon $|N,0\rangle$ is then identified as the lowest energy
solution with the associated quantum numbers, while the Roper
is obtained as the next state ($|N,1\rangle$) in the same spin --
isospin channel. Turning to the quantum numbers of the $\Lambda$
provides not only the energy $E_{\Lambda,0}$ and wave--function 
$\langle A,x|\Lambda,0\rangle$ of this hyperon but also the
analogous quantities for the radially excited $\Lambda$'s:
$E_{\Lambda,m}$ and $\langle A,x|\Lambda,m\rangle$ with $m\ge1$. These
calculations are repeated for the other spin -- isospin channels
yielding the spectrum not only of the ground state $\frac{1}{2}^+$
and $\frac{3}{2}^+$ baryons but also their monopole excitations. 
Of course, flavor symmetry breaking couples all possible $SU(3)$ 
representations as $\langle\mu|D_{88}|\mu^\prime\rangle$ is not
diagonal in $\mu$ and $\mu^\prime$. It also mixes various 
monopole excitations of the basis states $|\mu,n_{\mu}\rangle$.
When diagonalizing (\ref{hammatr})
we consider the basis built by the representations 
$\mbox{\boldmath $8$}$, $\overline{\mbox{\boldmath $10$}}$,
$\mbox{\boldmath $27$}$, $\overline{\mbox{\boldmath $35$}}$,
$\mbox{\boldmath $64$}$, $\overline{\mbox{\boldmath $81$}}$,
$\mbox{\boldmath $125$}$, $\overline{\mbox{\boldmath $154$}}$ for
the $\frac{1}{2}^+$ baryons and
$\mbox{\boldmath $10$}$, ${\mbox{\boldmath $27$}}$,
$\mbox{\boldmath $35$}$, $\overline{\mbox{\boldmath $35$}}$,
$\mbox{\boldmath $64$}$, $\overline{\mbox{\boldmath $28$}}$,
$\mbox{\boldmath $81$}$, $\overline{\mbox{\boldmath $81$}}$
$\mbox{\boldmath $125$}$, $\overline{\mbox{\boldmath $80$}}$
$\mbox{\boldmath $154$}$, $\overline{\mbox{\boldmath $254$}}$ for
the $\frac{3}{2}^+$ baryons. For the breathing degree of freedom we
include basis states which are up to $4{\rm GeV}$ above the ground
states of the flavor symmetric piece (\ref{freebreath}), {\it i.e.} 
$|\mbox{\boldmath $8$},0\rangle$ and $|\mbox{\boldmath $10$},0\rangle$ 
for the $\frac{1}{2}^+$ and $\frac{3}{2}^+$ baryons, respectively. This 
seems to be sufficient to get acceptable convergence when diagonalizing 
(\ref{hammatr}). It should be noted that not all of the above SU(3) 
representations appear in each isospin channel. For example, there 
are no $\Lambda$ and $\Xi$--type states in the antidecuplet,
$\overline{\mbox{\boldmath $10$}}$. 

\section{Results}

We expect three categories of model results. In the first place we 
have the ordinary low--lying $J=\frac{1}{2}$ and $J=\frac{3}{2}$ 
baryons together with their monopole excitations. Without flavor 
symmetry breaking these would be pure octet and decuplet states. 
In the second place we have $J=\frac{1}{2}$ states that are
dominantly members of the antidecuplet. Those that are non--exotic
mix with octet baryons and their monopole excitations. We are
particularly interested in dominantly antidecuplet baryons
($\Theta^+$ and $\Xi_{3/2}$) that do not have partners 
with identical quantum numbers in the octet. These antidecuplet baryons
cannot be constructed as three quark composites and are purely exotic. 
Third, we have $J=\frac{3}{2}$ baryons that originate from the 
$\mathbf{27}$--plet, {\it cf.} right panel of figure~\ref{fig_1}. 
Not all of them are exotic, as the 
$\Delta_{27},\Sigma^*_{27}$ and $\Xi^*_{27}$ have partners
in the decuplet. Finally there are also  $J=\frac{1}{2}$ baryons
in the $\mathbf{27}$--plet. They are heavier than the
$J=\frac{3}{2}$ members of the $\mathbf{27}$--plet and will 
therefore not be studied here. Note that we will consider
only mass differences (with respect to the nucleon) because
they are expected to be more reliably predicted in
chiral soliton models than absolute masses~\cite{Mo93}.

\subsection{Ordinary Baryons and their Monopole Excitations}

In table \ref{tab_1} the predictions for the mass differences with 
respect to the nucleon of the eigenstates of the full 
Hamiltonian~(\ref{hammatr}) are shown for two values 
of the Skyrme parameter $e$. These results were already reported
earlier~\cite{Sch91,Sch91a,We98}. For completeness and later comparison 
with the interesting $\mathbf{8}\oplus\overline{\mathbf{10}}$
picture of refs.~\cite{Ja03,Di03} we also discuss those results here.

\begin{table}
\caption{\label{tab_1}
The mass differences with respect to the nucleon 
of the eigenstates of the Hamiltonian (\protect\ref{hammatr}). 
The notation for the states appearing in this table is defined in
eq~(\protect\ref{bsbreath}). All numbers are in MeV.  Experimental 
data are taken from~\protect\cite{PDG02}, if available. Unless
otherwise noted, the data from~\protect\cite{PDG02} refer to four 
and three star resonances.  For the Roper resonance [$N(1440)$] we 
list both the Breit--Wigner~(BW) mass and the pole position~(PP) 
estimate of ref.~\protect\cite{PDG02}. The experimental states 
furnished with~"?" in the $|\Xi,1\rangle$ column are potential 
isospin $\frac{1}{2}$ candidates whose spin--party quantum numbers 
are not yet determined~\protect\cite{PDG02}.}
~
\newline
\centerline{
\begin{tabular}{c | c c c | c c c | c c c}
B& \multicolumn{3}{c|}{$m=0$} & \multicolumn{3}{c|}{$m=1$}
& \multicolumn{3}{c}{$m=2$} \\
\hline
& ~~$e$=5.0~~ & ~~$e$=5.5~~ & expt. 
& ~~$e$=5.0~~ & ~~$e$=5.5~~ & expt.
& ~~$e$=5.0~~ & ~~$e$=5.5~~ & expt. \\ 
\hline
\vspace{-0.3cm}
&&&&&&&&\\
N & \multicolumn{3}{c|}{Input}
            & 413 & 445 & 
\parbox[t]{2cm}{\vskip-0.4cm\large${\rm 501~BW}\atop{\rm 426~PP}$}
  & 836 & 869 & 771  \\
$\Lambda$ 
& 175 & 173 & 177 & 657 & 688 & 661 & 1081 & 1129 & 871 \\
\vspace{-0.3cm}
&&&&&&&&\\
$\Sigma$
& 284 & 284 & 254 & 694 & 722 & 721 & 1068 & 1096 & 
\parbox[t]{0.9cm}{\vskip-0.35cm
\large${\rm 831~(*)~}\atop{\rm 941~(**)}$}
\\
\vspace{-0.3cm}
&&&&&&&&\\
$\Xi$
& 382 & 380 & 379 & 941 & 971 & 
\parbox[t]{0.9cm}{\vskip-0.35cm\large${\rm 751}\atop{\rm 1011}$}(?)
& 1515 & 1324 & --- \\
\vspace{-0.3cm}
&&&&&&&&\\
\hline
$\Delta$
& 258 & 276 & 293 & 640 & 680 & 661 & 974 & 1010 & 981 \\
$\Sigma^*$
& 445 & 460 & 446 & 841 & 878 & 901 & 1112 & 1148 & 1141 \\
$\Xi^*$
& 604 & 617 & 591 & 1036 & 1068 & --- & 1232 & 1269 & --- \\
$\Omega$
& 730 & 745 & 733 & 1343 & 1386 & --- & 1663 & 1719 & --- \\
\end{tabular}}
\end{table}
The agreement with the experimental data is quite astonishing, not
only for the ground state but also for the radial excitations. Only 
the prediction for the Roper resonance ($|N,1\rangle$) is a bit on the 
low side when compared to the empirical Breit--Wigner mass but 
agrees reasonably well with the estimated pole position of that
resonance. This is common for the breathing mode approach in soliton
models~\cite{Ha84,HH84}.
For other nucleon (and $\Delta$) resonances the discrepancy between
the Breit--Wigner mass and the pole position is much smaller, of
the order of $20{\rm MeV}$ or less~\cite{PDG02}. Thus there is no 
need to distinguish between them. As far as data are 
available and the quantum numbers are definite, the other first 
excited states are quite well reproduced. For the $\frac{1}{2}^+$ 
baryons the energy eigenvalues for the second excitations overestimate 
the corresponding empirical data somewhat. However, the pattern 
$M(|N,2\rangle)<M(|\Sigma,2\rangle)<M(|\Lambda,2\rangle)$ is reproduced
if $|\Sigma,2\rangle$ is identified with the single star resonance
$\Sigma(1770)$ that is about $830{\rm MeV}$ heavier than the nucleon. 
The predicted $\Sigma$ and $\Lambda$ type states 
with $m=2$ are about $100$--$200{\rm MeV}$ too high. It is worth 
noting that in the nucleon channel the $m=3$ state is predicted to 
only be about $40{\rm MeV}$ higher than the $m=2$ state, {\it i.e.} 
still within the regime where the model is assumed to be applicable. 
This is interesting because empirically it is suggestive that there 
might exist more than only one resonance in the concerned energy
region~\cite{Ba95}. For the $\frac{3}{2}^+$ baryons with $m=2$ the 
agreement with data is even better, on the 3\% level. The particle 
data group~\cite{PDG02} lists two ``three star'' isospin--$\frac{1}{2}$ 
$\Xi$ resonances whose spin--parity quantum numbers are not yet 
determined at $751$ and $1011{\rm MeV}$ above the nucleon. Turning
to absolute masses these are $\Xi(1690)$ and $\Xi(1950)$, 
respectively. The present 
model suggests that the latter is indeed $J^\pi=\frac{1}{2}^+$, while 
the former seems to belong to a different channel. For example, the 
sizable mass difference between the lowest lying ($m=0$) $\Sigma$ and 
$\Xi$ motivates the speculation that the $\Xi(1690)$ should be 
considered as a vibrational ($\hbar\omega$) excitation of the octet 
$\Xi$ and that it originates from a multiplet with $N(1520)$, 
$\Lambda(1520)$ and $\Sigma(1580)$. These are $D$--wave resonances 
with $J^\pi=\frac{3}{2}^-$ that are prominently observed in scattering 
calculations off the soliton~\cite{Schw89}. On the other hand the 
established $D$--wave $J^\pi=\frac{3}{2}^-$ state $\Xi(1820)$ may be 
associated with a multiplet formed with $N(1700)$, $\Lambda(1690)$, 
$\Sigma(1670)$. Such a picture is somewhat appealing as both
$J^\pi=\frac{3}{2}^-$ octets would have (almost) degenerate nucleon
and $\Lambda$ states. Here we finally note that the model state
$|\Xi,1\rangle$ is dominated by the first radial excitation
in the octet and the ground state in the $\mathbf{27}$--plet,
as ${C_{\mathbf{8},1}^{(\Xi,1)}}^2\approx 
{C_{\mathbf{27},0}^{(\Xi,1)}}^2\approx0.4$. 

On the whole, the present model gives fair agreement with the 
available data. This certainly supports the picture of coupled 
monopole and rotational modes. The important message is that three 
flavor chiral models do \emph{not} predict any novel states in the
energy regime between $1$ and $2{\rm GeV}$ in the non--exotic channels 
as a consequence of including higher dimensional $SU_F(3)$ flavor 
representations.

\subsection{Exotic Baryons from the Antidecuplet}

The $\overline{\mathbf{10}}$ representation contains two states 
that possess quantum numbers that cannot be constructed as three
quark composites, the $\Theta^+$ and the $\Xi_{3/2}$. The model 
prediction for these states are listed in table~\ref{tab_2} and 
compared to available data~\cite{Thetaexp,NA49}. As for the 
non--exotic baryons, we have computed the respective mass differences 
to the nucleon and added the experimental nucleon mass to set
the overall mass scale.  We also compare to 
a chiral soliton model calculation~\cite{Wa03} that does not include 
a dynamical treatment of the monopole excitation. In that calculation 
parameters have been tuned to reproduce the mass of the lightest
exotic pentaquark, $\Theta^+$. The inclusion of the monopole
excitation increases the mass of the $\Xi_{3/2}$ slightly
and brings it closer to the empirical value. We furthermore note 
that the first prediction~\cite{Di97} for the mass of the $\Xi_{3/2}$ 
that was based on identifying $N(1710)$ with the nucleon like state 
in the antidecuplet resulted in a far too large mass of $2070{\rm MeV}$. 
There are other chiral soliton model studies of the pentaquarks of the 
antidecuplet. However, those either take $M_{\Xi_{3/2}}$ as 
input~\cite{Bo03,Di03}, adopt the assumptions of ref.~\cite{Di97} or 
are less predictive because the model parameters are allowed to 
vary considerably~\cite{El04}. In any event, it is desirable to 
independently confirm the NA49 analysis that led to 
$M_{\Xi_{3/2}}=1.862\pm0.002$.

\begin{table}
\caption{\label{tab_2}
Masses of the eigenstates of the Hamiltonian (\protect\ref{hammatr}) 
the exotic baryons $\Theta^+$ and $\Xi_{3/2}$. 
Experimental data are the average of refs.~\protect\cite{Thetaexp} 
for $\Theta^+$ and the NA49 result for $\Xi_{3/2}$~\cite{NA49}.
We also compare the predictions for the ground state ($m=0$) to the 
treatment of ref.~\cite{Wa03} that does not quantize the
monopole mode. All energies are in~GeV.}
~
\newline
\centerline{
\begin{tabular}{c | c c c |c | c c c}
B& \multicolumn{4}{c|}{$m=0$} & \multicolumn{3}{c}{$m=1$}\\
\hline
& ~~$e$=5.0~~ & ~~$e$=5.5~~ & expt. & WK~\cite{Wa03}
& ~~$e$=5.0~~ & ~~$e$=5.5~~ & expt. \\
\hline
$\Theta^+$ & $1.57$ & $1.59$ & $1.537\pm0.010$ & $1.54$ &
$2.02$ & $2.07$ & -- \\
$\Xi_{3/2}$ & $1.89$ & $1.91$ & $1.862\pm0.002$ & $1.78$ &
$2.29$ & $2.33$ & -- 
\end{tabular}}
\end{table}
In all cases where comparison with data is possible, we observe that 
without any fine--tuning the model prediction is only about 
$30$--$50{\rm MeV}$ higher than the data. In view of the approximative 
nature of the model this level of agreement should be viewed as good. 
In particular the mass difference between the two potentially observed 
exotics is reproduced within $10{\rm MeV}$, {\it cf.} table~\ref{tab_2}.
Notably, both the empirical data as well as the present model
calculation for this mass difference are slightly above the upper
limit $M_{\Xi_{3/2}}-M_{\Theta^+}<299{\rm MeV}$ obtained in a 
quark model with a color magnetic interaction~\cite{Ka04}.

\subsection{Baryons from the 27--plet}

The $\mathbf{27}$ dimensional representation allows for both, baryons
with spin $J=\frac{1}{2}$ and $J=\frac{3}{2}$. Since 
$\alpha(x)>\beta(x)$~\cite{Sch91,Sch91a} the $J=\frac{3}{2}$
baryons will be lighter according to the 
Hamiltonian~(\ref{freebreath},\ref{hammatr}).
We will therefore only consider the $J=\frac{3}{2}$ baryons.
As is shown in figure~\ref{fig_1}, the $\mathbf{27}$--plet
contains states with the quantum numbers of the baryons
that are also contained in the decuplet of the low--lying
$J=\frac{3}{2}$ baryons: $\Delta,\Sigma^*$ and $\Xi^*$. 
Under flavor symmetry breaking 
these states of the decuplet and $\mathbf{27}$--plet mix with 
each other as well as with the respective monopole excitations. 
Stated otherwise, $\Delta,\Sigma^*$ and $\Xi^*$ states that 
emerge from the $\mathbf{27}$--plet have been already considered
when diagonalizing the full Hamiltonian~(\ref{hammatr}) for the 
states with quantum numbers of the low--lying $J=\frac{3}{2}$ 
baryons. Eventually these members of the $\mathbf{27}$--plet 
represent the dominant amplitude in the exact eigenstates 
$|\Delta,1\rangle$ and/or $|\Delta,2\rangle$ and similarly for 
$\Sigma^*$ and~$\Xi^*$. The corresponding eigenvalues are displayed 
and compared to available data in table~\ref{tab_1}.
In table~\ref{tab_3} we present the model predictions for
the $J=\frac{3}{2}$ baryons that emerge from the $\mathbf{27}$--plet
but do not have partners in the lower dimensional representations.
Again, we have used the experimental nucleon mass to set the
overall mass scale.
\begin{table}
\caption{\label{tab_3}
Predicted masses 
of the eigenstates of the Hamiltonian (\protect\ref{hammatr}) for
the exotic $J=\frac{3}{2}$ baryons with $m=0$ and $m=1$ that originate
from the $\mathbf{27}$--plet. The corresponding hypercharge (Y)
and isospin (I) quantum numbers are listed. We also compare the
$m=0$ case to treatments without breathing mode quantization of 
refs.~\cite{Wa03,Bo03,Wu03}. The results of ref.~\cite{Bo03} 
have been confirmed in ref.~\cite{El04}. All numbers are in GeV.}
~
\newline
\centerline{
\begin{tabular}{c | c c | c c |c|c|c|cc}
\multicolumn{3}{c|}{B}& \multicolumn{5}{c|}{$m=0$} 
& \multicolumn{2}{c}{$m=1$}\\
\hline
& ~Y~ & ~I~ & ~~$e$=5.0~~ & ~~$e$=5.5~~ 
& WK~\cite{Wa03} & BFK~\cite{Bo03} & WM~\cite{Wu03} 
&~~$e$=5.0~~ & ~~$e$=5.5~~ \\
\hline
$\Theta_{27}$ &$2$ & $1$ & $1.66$ & $1.69$ & $1.67$ 
& $1.60$ & $1.60$ & 2.10 & 2.14\\
$N_{27}$ &$1$ & $1/2$ & $1.82$ & $1.84$ & $1.76$ 
& $ -- $ & $1.73$ & 2.28 & 2.33  \\
$\Lambda_{27}$ &$0$ & $0$ & $1.95$ & $1.98$ & $1.86$ 
& $ -- $ & $1.86$ & 2.50 & 2.56\\
$\Gamma_{27}$ &$0$ & $2$ & $1.70$ & $1.73$ & $1.70$ 
&  1.70  & $1.68$ & 2.12 & 2.17 \\
$\Pi_{27}$ &$-1$ & $3/2$ & $1.90$ & $1.92$ & $1.84$ 
& $1.88$ & $1.87$ & 2.35 & 2.40\\
$\Omega_{27}$ &$-2$ & $1$ & $2.08$ & $2.10$ & $1.99$ 
& $2.06$ & $2.07$ & 2.54 & 2.59
\end{tabular}}
\end{table}
Essentially the masses displayed in table~\ref{tab_3} are model 
predictions for resonances that are yet to be observed. Nevertheless 
we note that the particle data group~\cite{PDG02} lists two states 
with the quantum numbers of $N_{27}$ and $\Lambda_{27}$ at $1.72$ 
and $1.89{\rm GeV}$, respectively, that fit reasonably well into the 
model calculation. In all channels the $m=1$ states turn out to be 
about $500{\rm MeV}$ heavier than the exotic ground states.
When combined with the first excited states of $\Delta$, $\Sigma^*$, 
and $\Xi^*$ that are listed under $m=1$ in
table~\ref{tab_1} it is interesting to observe
that the masses of states that are degenerate in
hypercharge decrease with isospin, that is
$M_{|\Delta,1\rangle}<M_{|N_{27},0\rangle}$,
$M_{|\Gamma_{27},0\rangle}<M_{|\Sigma^*,1\rangle}
<M_{|\Lambda_{27},0\rangle}$,
$M_{|\Pi_{27},0\rangle}<M_{|\Xi^*,1\rangle}$, and
$M_{|\Omega_{27},0\rangle}<M_{|\Omega,1\rangle}$.

The comparison with treatments~\cite{Wa03,Bo03,Wu03,El04} that do not
quantize the monopole degree of freedom indicates that for the 
exotic pentaquark states that do not have partners in lower dimensional
$SU_F(3)$ representations the mixing of rotational and monopole 
modes is a minor effect and can eventually be discarded. The 
treatment of ref.~\cite{Wa03} is similar to the present one in 
the sense that flavor symmetry breaking is treated to all orders 
when diagonalizing the Hamiltonian for the collective degrees of freedom. 
On the other hand, in refs.~\cite{Bo03,Wu03,El04} a first order approximation
has been adopted. Not surprisingly, the model calculations that do not 
quantize the monopole excitations find masses for the non--exotic states 
$\Delta_{27}$, $\Sigma^*_{27}$ and $\Xi^*_{27}$, that lie in between
the $m=1$ and $m=2$ eigenvalues of $\Delta$, $\Sigma^*$ and $\Xi^*$,
{\it cf.} table~\ref{tab_1}. For example, the authors of ref.~\cite{Wa03} 
predict $M_{\Delta_{27}}\approx1.78$, $M_{\Sigma^*_{27}}\approx1.86$
and $M_{\Xi^*_{27}}\approx1.95$. Only for $e=5.5$ both, the
$m=1$ ($1.07+0.94=2.01{\rm GeV}$) and the $m=2$ 
($1.27+0.94=2.21{\rm GeV}$) for $\Xi^*$ 
are slightly above that prediction for $\Xi^*_{27}$. This may be
connected to the observation that we presently also find a
somewhat heavier $\Xi_{3/2}$ than in ref.~\cite{Wa03}, 
{\it cf.} table~\ref{tab_2}.

The statement that the dynamical treatment of the monopole
degree of freedom does not significantly alter the mass eigenvalues
of the exotic baryons is also supported by
the observation that those states acquire their dominant
contributions to the wave--functions from the 
$\mathbf{27}$--plet.  For a convenient discussion thereof
we present in table~\ref{tab_4} the sums
\be
P_\mu=\sum_{n_\mu}\left[C_{\mu,n_\mu}^{(B,m)}\right]^2
\label{sumsq}
\ee
of the squared amplitudes that are defined in eq.~(\ref{bsbreath}).
The $P_\mu$ are to be interpreted as the probability to find
a state in a given irreducible representation, $\mu$, of $SU_F(3)$.
\begin{table}
\caption{\label{tab_4}
Mixing pattern for the $J^\pi=\frac{3}{2}^+$ states. In case of 
$\Delta$, $\Sigma^*$, $\Gamma_{27}$, and $\Pi_{27}$ the notations
$\mathbf{35}_1$ and $\mathbf{35}_2$ denote unspecified 
orthonormal linear combinations of the corresponding states from 
$\mathbf{35}$ and $\overline{\mathbf{35}}$. In all other cases 
$\mathbf{35}_1$ and $\mathbf{35}_2$ equal $\mathbf{35}$ 
and $\overline{\mathbf{35}}$, respectively.
Listed are the probabilities $P_\mu$ defined
in eq.~(\ref{sumsq}) for $e=5.0$.}
~
\newline
\centerline{
\begin{tabular}{c | c c c c c}
$B,m$& ~$\mu=\mathbf{10}$~ & ~$\mu=\mathbf{27}$~ &
~$\mu=\mathbf{35}_1$~ & ~$\mu=\mathbf{35}_2$~ &
~$\mu=\mathbf{64}$~\\
\hline
$|\Delta,0\rangle$ & 0.60 & 0.30 & 0.04 & 0.03 & 0.02 \\
$|\Delta,1\rangle$ & 0.54 & 0.22 & 0.06 & 0.09 & 0.05 \\
$|\Delta,2\rangle$ & 0.19 & 0.61 & 0.06 & 0.09 & 0.03 \\
$|\Delta,3\rangle$ & 0.57 & 0.10 & 0.04 & 0.16 & 0.04 \\
$|\Sigma^*,0\rangle$ & 0.73 & 0.18 & 0.06 & 0.01 & 0.01 \\
$|\Sigma^*,1\rangle$ & 0.47 & 0.27 & 0.11 & 0.05 & 0.07 \\
$|\Sigma^*,2\rangle$ & 0.14 & 0.67 & 0.11 & 0.03 & 0.02 \\
$|\Xi^*,0\rangle$ & 0.87 & 0.07 & 0.05 &  --  & 0.01 \\
$|\Xi^*,1\rangle$ & 0.31 & 0.47 & 0.13 &  --  & 0.05 \\
$|\Xi^*,2\rangle$ & 0.23 & 0.52 & 0.19 &  --  & 0.01 \\
$|\Omega,0\rangle$ & 0.97 &  --  & 0.03 &  --  & 0.00 \\
$|\Omega,1\rangle$ & 0.55 &  --  & 0.39 &  --  & 0.06 \\
$|\Omega,2\rangle$ & 0.50 &  --  & 0.36 &  --  & 0.12 \\
\hline
$|\Theta_{27},0\rangle$ & --  & 0.76 & --  & 0.16 & 0.05 \\
$|\Theta_{27},1\rangle$ & --  & 0.54 & --  & 0.25 & 0.08 \\
$|N_{27},0\rangle$ & --  & 0.87 & --  & 0.07 & 0.06 \\
$|N_{27},1\rangle$ & --  & 0.49 & --  & 0.20 & 0.19 \\
$|\Lambda_{27},0\rangle$ & --  & 0.93 & --  &  --  & 0.06 \\
$|\Lambda_{27},1\rangle$ & --  & 0.56 & --  &  --  & 0.37 \\
$|\Gamma_{27},0\rangle$ & --  & 0.73 & 0.09 & 0.10 & 0.06 \\
$|\Gamma_{27},1\rangle$ & --  & 0.52 & 0.14 & 0.11 & 0.37 \\
$|\Pi_{27},0\rangle$ & --  & 0.82 & 0.11 & 0.00 & 0.05 \\
$|\Pi_{27},1\rangle$ & --  & 0.46 & 0.25 & 0.00 & 0.19 \\
$|\Omega_{27},0\rangle$ & --  & 0.90 & --  & 0.06 & 0.04 \\
$|\Omega_{27},1\rangle$ & --  & 0.29 & --  & 0.41 & 0.16
\end{tabular}}
\end{table}
As is already known for the $J^\pi=\frac{1}{2}^+$ 
states~\cite{Sch91,Sch91a,We98,We00}, the lowest lying state
in a channel with given hypercharge and isospin is strongly
dominated by the state from the lowest allowed $SU_F(3)$
representation, {\it e.g.} the nucleon state is dominantly
$|\mathbf{8},0\rangle$.  On the other hand, the first 
excited state turns out to be a complicated linear combination 
of all possible basis states, again a pattern also observed for
the $J=\frac{1}{2}$ eigenstates of the Hamiltonian~(\ref{hammatr}).
In the following section we will discuss the mixing pattern for 
the $J=\frac{1}{2}$ states in more detail.

\section{Comparison with the $\mathbf{8}\oplus\overline{\mathbf{10}}$
picture}

\begin{figure}
\setlength{\unitlength}{1.0mm}
\begin{picture}(150,120)
\thicklines
\put(-7,82){E[GeV]}
\put(-7,29){1.5}
\put(-1,30){\line(1,0){2}}
\put(-7,59){1.8}
\put(-1,60){\line(1,0){2}}
\put(0,10){\line(0,1){65}}
\put(0,75){\vector(0,1){5}}
\put(15,90){$e=5.0$}
\put(10,15.3){\line(1,0){20}}
\put(35,14.3){$|N,1\rangle$}
\put(10,39.7){\line(1,0){20}}
\put(35,38.4){$|\Lambda,1\rangle$}
\put(10,43.4){\line(1,0){20}}
\put(35,42.5){$|\Sigma,1\rangle$}
\put(10,68.0){\line(1,0){20}}
\put(35,65.0){$|\Xi,1\rangle$}
\put(10,57.7){\line(1,0){20}}
\put(35,56.4){$|N,2\rangle$}
\put(10,80.7){\line(1,0){20}}
\put(35,78.8){$|\Sigma,2\rangle$}
\put(10,37.0){\line(1,0){20}}
\put(35,35.0){$|\Theta^+,0\rangle$}
\put(10,69.0){\line(1,0){20}}
\put(35,69.5){$|\Xi_{3/2},0\rangle$}
\multiput(10.2,82.1)(2,0){10}{\line(1,0){1}}
\put(35,83.8){$|\Lambda,2\rangle$}
\multiput(10.2,61.1)(2,0){10}{\line(1,0){1}}
\put(35,60.8){$|N,3\rangle$}
\put(65,90){JW~\cite{Ja03}}
\put(60,34){\line(1,0){20}}
\put(85,33){$\Theta^+$}
\put(60,23){\line(1,0){20}}
\put(85,22){$N$}
\put(60,42){\line(1,0){20}}
\put(85,41){$\Lambda,\Sigma$}
\put(60,58){\line(1,0){20}}
\put(85,57){$N_S$}
\put(60,77){\line(1,0){20}}
\put(85,76){$\Sigma_S$}
\put(60,65){\line(1,0){20}}
\put(85,64){$\Xi,\Xi_{3/2}$}
\put(122,90){DP~\cite{Di03}}
\put(112,82){A}
\put(105,27){\line(1,0){15}}
\put(122,26){$N_1$}
\put(105,45.6){\line(1,0){15}}
\put(122,45.0){$N_2$}
\put(105,37){\line(1,0){15}}
\put(122,36){$\Lambda$}
\put(105,43.5){\line(1,0){15}}
\put(122,42.5){$\Sigma_1$}
\put(105,57.8){\line(1,0){15}}
\put(122,56.8){$\Sigma_2$}
\put(105,50){\line(1,0){15}}
\put(122,49){$\Xi$}
\put(105,66.2){\line(1,0){15}}
\put(122,65.2){$\Xi_{3/2}$}
\put(105,33.9){\line(1,0){15}}
\put(122,32.9){$\Theta^+$}
\put(140,82){B}
\put(135,26){\line(1,0){15}}
\put(152,25){$N_1$}
\put(135,47.6){\line(1,0){15}}
\put(152,46.6){$N_2$}
\put(135,37.5){\line(1,0){15}}
\put(152,36.5){$\Lambda$}
\put(135,46){\line(1,0){15}}
\put(152,44){$\Sigma_1$}
\put(135,59.9){\line(1,0){15}}
\put(152,58.9){$\Sigma_2$}
\put(135,51){\line(1,0){15}}
\put(152,50){$\Xi$}
\put(135,66.2){\line(1,0){15}}
\put(152,65.2){$\Xi_{3/2}$}
\put(135,33.9){\line(1,0){15}}
\put(152,32.9){$\Theta^+$}
\end{picture}
\caption{\label{fig_2}Comparison of the predicted spectra in various 
models for pentaquarks with $J=\frac{1}{2}$. The present model is 
labeled by $e=5.0$, the parameter chosen for presentation.
Two additional states that fall into the depicted energy range 
are indicated by dotted lines. For the non--exotic states the model 
results are taken from table~\ref{tab_1} with the physical nucleon
mass added. JW denotes a calculation within the 
Jaffe--Wilczek diquark model~\cite{Ja03} with the parameters
$M_0=1.44{\rm GeV}$ $m_s=0.11{\rm GeV}$ and $\alpha=0.06{\rm GeV}$
substituted into their mass formula. In columns A and B the
results of the two scenarios of the Diakonov--Petrov 
approach~\cite{Di03} are shown. We have adopted the 
notation used in the respective papers.}
\end{figure}
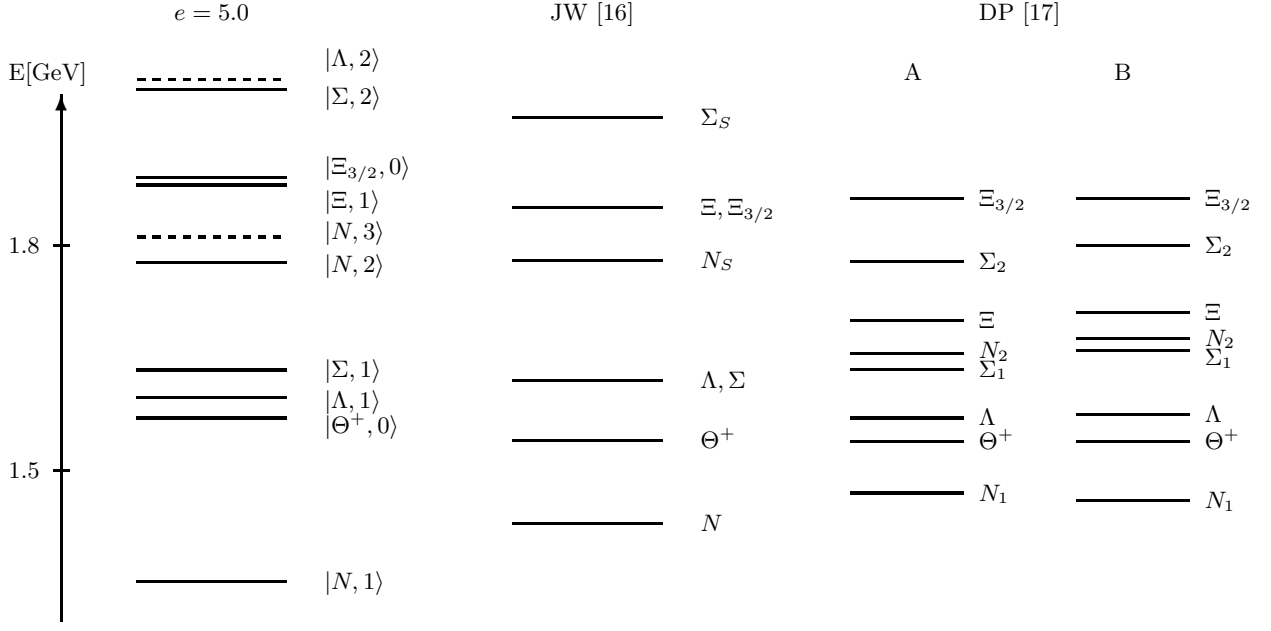

In the present approach to compute the masses of pentaquarks there exists
an important and obvious interplay between the radially excited octet 
baryons and the ground state baryons in the antidecuplet. We expect 
significant mixing for states in the $1.3$ to $1.8{\rm GeV}$ energy 
regime because the eigen--energies ${\cal E}_{\mathbf{8},1}$ and 
${\cal E}_{\overline{\mathbf{10}},0}$ of the basis states are not 
only about $0.5{\rm GeV}$ above ${\cal E}_{\mathbf{8},0}$ but also 
quite close together. For example ${\cal E}_{\mathbf{8},1}-
{\cal E}_{\overline{\mathbf{10}},0}\approx18{\rm MeV}$ for $e=5.0$. 
Hence, even if the associated symmetry breaking matrix elements
$\langle\mu,n_{\mu}|sD_{88}|\mu^\prime,n_{\mu^\prime}\rangle$
in eq.~(\ref{hammatr}) were small we expect large mixing effects.
The eigen--energies of other basis states are several hundred MeV 
away and thus it is plausible that the $J^\pi=\frac{1}{2}^+$ 
baryons in the energy regime between $1.3$ and $1.8{\rm GeV}$ may 
be viewed as members of the direct sum 
$\mathbf{8}\oplus\overline{\mathbf{10}}$ of $SU_F(3)$ representations, 
at least approximately. In such a
$\mathbf{8}\oplus\overline{\mathbf{10}}$ scenario the 
$\Lambda$ and $\Xi$ resonances are pure octet while $\Theta^+$ and 
$\Xi_{3/2}$ are pure antidecuplet. Mixing can only occur for nucleon 
and $\Sigma$ type states. There are two approaches to arrange the 
excited baryons within such a direct sum. In ref.~\cite{Ja03} a diquark 
picture has been adopted that leads to an ideal mixing between 
baryons of identical quantum numbers within the $\mathbf{8}$ and 
$\overline{\mathbf{10}}$ representations such that the eigenstates
have minimal or maximal strangeness content. For example, the octet 
nucleon and the antidecuplet nucleon ($N^\prime$ in figure~\ref{fig_1})
mix to built eigenstates with the quark structure $uud(\bar{u}u)$ and
$uud(\bar{s}s)$, modulo the isospin partners. Then the baryon mass 
formula essentially counts the number of strange quarks and antiquarks 
that are contained in the resonance. In ref.~\cite{Di03} the 
$\mathbf{8}\oplus\overline{\mathbf{10}}$ decomposition was taken as 
starting point. For these octet and antidecuplet states the pattern of 
flavor symmetry breaking was adopted that results from a first order 
calculation in soliton models. Those authors additionally assumed that 
$\Xi(1690)$ -- whose spin--parity quantum numbers are yet to be 
determined~\cite{PDG02} -- was the octet partner of $\Lambda(1600)$ to 
estimate the octet mass parameters and fixed the antidecuplet mass 
parameters from $\Theta^+(1535)$ and $\Xi_{3/2}(1862)$. With that
input they computed the mixing angle between the $\mathbf{8}$ and
$\overline{\mathbf{10}}$ representations and predicted so far
unobserved nucleon and $\Sigma$ resonances in the $1650-1810{\rm MeV}$
region. In figure~\ref{fig_2} we compare the mass spectra of these 
two $\mathbf{8}\oplus\overline{\mathbf{10}}$ scenarios with the
present model calculation.
Surprisingly the spectra obtained in the current model calculation 
and that of the Jaffe--Wilczek ideal mixing scenario are very similar, 
at least qualitatively. The most apparent similarity is the (almost) 
degeneracy of $\Xi_{3/2}$ and the first excited $\Xi$.  
The model of ref.~\cite{Ja03} also has 
degenerate $\Lambda$ and $\Sigma$ states. In the present approach such 
a degeneracy is also indicated but not very pronounced. Furthermore both
treatments yield the second excited $\Sigma$ way above the $\Xi_{3/2}$ 
as well as a large gap between the first exited $\Sigma$ and the 
second exited nucleon. This is somewhat different in the analysis of 
ref.~\cite{Di03}: Most obvious is the large gap between $\Xi$ and 
$\Xi_{3/2}$ with the second excited $\Sigma$ sitting in between. For that
result it was crucial to assume that $\Xi(1690)$ is in the same 
$SU_F(3)$ multiplet as the $\Lambda(1600)$. While $\Lambda(1600)$ 
has $J^\pi=\frac{1}{2}^+$, the spin--parity quantum numbers
of $\Xi(1690)$ are not yet determined experimentally.
If indeed $\Xi(1690)$ were the partner of the
$\Lambda(1600)$ one would expect to also obtain such a state
in the $J^\pi=\frac{1}{2}^+$ channel in the quark model
calculation of ref.~\cite{Gl97}. However, in that
calculation the first excited $\Xi$ with $J^\pi=\frac{1}{2}^+$
shows up significantly higher, at about $1850{\rm MeV}$,
{\it i.e.} similar to the present prediction for
$|\Xi,1\rangle$, {\it cf.} table~\ref{tab_1}. As already
discussed in section IIIC, it seems more plausible to 
assign $J^\pi=\frac{3}{2}^-$ to $\Xi(1690)$ and 
$J^\pi=\frac{1}{2}^+$ to $\Xi(1950)$. From tables~\ref{tab_3} 
and~\ref{tab_4} we recognize a similarity between the structure 
of the nucleon and $\Delta$ states. The first excited state resides
mostly in the lowest possible dimensional $SU_F(3)$ 
representation, {\it i.e.} a monopole excitation of the
ground state. The $m=2$ state is dominated by the member
of the next--to--lowest possible dimensional $SU_F(3)$
representation while $m=2$ again is in the same multiplet 
as the ground and originates from the second monopole
excitation.

Although the spectrum in the present model turns out
to be similar to the Jaffe--Wilczek scenario, the
mixing pattern is considerably more complicated than
an ideal direct sum $\mathbf{8}\oplus\overline{\mathbf{10}}$.
In table~\ref{tab_5} we give the mixing for the 
states depicted in figure~\ref{fig_2} in form 
of the  probabilities $P_\mu$ that have been defined in 
eq.~(\ref{sumsq}).
\begin{table}
\caption{\label{tab_5}
Mixing pattern for the low--lying $J^\pi=\frac{1}{2}^+$ 
baryons and the states shown in figure~\ref{fig_2}.
Listed are the probabilities $P_\mu$ defined 
in eq.~(\ref{sumsq}) for $e=5.0$.}
~
\newline
\centerline{
\begin{tabular}{c | c c c c c}
$B,m$& ~$\mu=\mathbf{8}$~ & ~$\mu=\overline{\mathbf{10}}$~ &
~$\mu=\mathbf{27}$~ & ~$\mu=\overline{\mathbf{35}}$~ &
~$\mu=\mathbf{64}$~\\
\hline
$|N,0\rangle$ & 0.87 & 0.06 & 0.05 & 0.01 & 0.00 \\
$|N,1\rangle$ & 0.59 & 0.15 & 0.16 & 0.05 & 0.03 \\
$|N,2\rangle$ & 0.12 & 0.68 & 0.10 & 0.05 & 0.04 \\
$|N,3\rangle$ & 0.57 & 0.17 & 0.08 & 0.12 & 0.03 \\
\hline
$|\Lambda,0\rangle $ & 0.93 & -- & 0.06 & -- & 0.00 \\
$|\Lambda,1\rangle $ & 0.58 & -- & 0.34 & -- & 0.07 \\
$|\Lambda,2\rangle $ & 0.58 & -- & 0.23 & -- & 0.17 \\
\hline
$|\Sigma,0\rangle$ & 0.88 & 0.08 & 0.04 & 0.01 & 0.00 \\
$|\Sigma,1\rangle$ & 0.37 & 0.33 & 0.15 & 0.10 & 0.03 \\
$|\Sigma,2\rangle$ & 0.11 & 0.69 & 0.13 & 0.04 & 0.04 \\
\hline
$|\Xi,0\rangle$ & 0.96 & -- & 0.04 & -- & 0.00 \\
$|\Xi,1\rangle$ & 0.49 & -- & 0.42 & -- & 0.07 \\
\hline
$|\Theta^+,0\rangle$ & -- & 0.85 & -- & 0.14 & -- \\
\hline
$|\Xi_{3/2},0\rangle$ & -- & 0.76 & 0.12 & 0.10 & 0.01
\end{tabular}}
\end{table}
In the $\mathbf{8}\oplus\overline{\mathbf{10}}$ scenarios 
of refs.~\cite{Ja03,Di03} both
the $|\Lambda,1\rangle$ and the $|\Xi,1\rangle$ would be 
pure octet states. In the present model calculation we 
find, however, that there is significant admixture of
the partners from the $\mathbf{27}$--plet, at the order
of 40\% in the squared amplitude, $P_\mu$. Furthermore the
states $|N,1\rangle$ and $|N,2\rangle$ as well as
$|\Sigma,1\rangle$ and $|\Sigma,2\rangle$ are not simple
linear combinations of the corresponding octet and antidecuplet 
states but also contain sizable contributions from their partners
in the $\mathbf{27}$--plet. In ref.~\cite{Ja03} the mixing 
pattern for $\mathbf{8}$ and $\overline{\mathbf{10}}$ states 
is ideal for the strangeness content, {\it i.e.} $N$ and 
$\Sigma$ have minimal strangeness content while it is maximal 
for $N_S$ and $\Sigma_S$. In order to further compare the 
present scenario we therefore estimate the strangeness
content as the matrix element~\cite{Sch91},
\be
S(B,m)=\frac{1}{3}\langle B,m|\left[1-D_{88}\right]|B,m\rangle
\label{stcont}
\ee
for the nucleon and $\Sigma$ eigenstates of the 
Hamiltonian~(\ref{hammatr}). In the soliton approach baryons naturally 
possess mesonic (quark--antiquark) clouds. It is
therefore quite instructive to first compare 
the model results to the flavor symmetric formulation. In that case 
the nucleon and $\Sigma$ ground states are pure octet and the 
above matrix elements are simple Clebsch--Gordan coefficients:
$S(N,\mathbf{8})=23\%$ and $S(\Sigma,\mathbf{8})=37\%$. Similarly 
the antidecuplet states have $S(N,\overline{\mathbf{10}})=29\%$ 
and $S(\Sigma,\overline{\mathbf{10}})=33\%$.
This large value for the nucleon reflects a large 
$\bar{s}s$ cloud that is easily excited when the strange quark
is assumed to be (almost) massless, as it is the case in the flavor 
symmetric treatment. In table~\ref{tab_6} we list the 
strangeness content for the three lowest nucleon and 
$\Sigma$ states.
\
\begin{table}
\caption{\label{tab_6}
Strangeness content in per cent as computed from 
eq.~(\ref{stcont}) for
$e=5.0$.}
~
\newline
\centerline{
\begin{tabular}{c | c c c |c c c}
& $|N,0\rangle$ & $|N,1\rangle$ & $|N,2\rangle$
& $|\Sigma,0\rangle$ & $|\Sigma,1\rangle$ & $|\Sigma,2\rangle$\\
\hline
$S$& 16 & 14 & 19 & 30 & 23 & 29
\end{tabular}}
\end{table}
As expected, the inclusion of flavor symmetry breaking,
{\it i.e.} the increase of the strange quark mass, leads to
an overall reduction. Furthermore
the lower lying excited states in the nucleon and $\Sigma$ 
channels possess smaller strangeness content than the 
second excited states. This is in the logic of the ideal
mixing pattern~\cite{Ja03} but not quite as pronounced as 
to maximize or minimize the strangeness content. 
Remarkably we find that the strangeness content of the first
excited nucleon and $\Sigma$ type states is even less than
that of the respective ground states.

\section{Conclusion}

We have analyzed the interplay between rotational and monopole excitations 
for the spectrum of pentaquarks in a chiral soliton model. The main issue 
in this approach has been the elevation of the scaling degree of 
freedom to a dynamical quantity which subsequently has been quantized 
canonically at the same footing as the (flavor) rotational modes.
In this manner not only the ground states in individual irreducible 
$SU_F(3)$ representations are eigenstates of the (flavor symmetric 
part of the) Hamiltonian but also all their radial excitations. It 
turns out that the various rotational and monopole excitations mix via 
the flavor symmetry breaking term in this Hamiltonian. In determining 
the baryon spectrum we have treated flavor symmetry breaking exactly 
rather then only at first order; an approximation often 
performed~\cite{Di97,Di03,Bo03,Wu03,El04}. Thus, even though the chiral 
soliton approach initiates from a flavor symmetric formulation, 
it is quite well capable of accounting for large deviations thereof. 

The spectrum of the low--lying $\frac{1}{2}^+$ and $\frac{3}{2}^+$ 
baryons is reasonably well reproduced. It is also important to note 
that the results for static properties such as magnetic moments and 
axial transition matrix elements that parameterize the amplitudes of 
semileptonic hyperon decays are in acceptable agreement with the 
empirical data~\cite{Sch91}. This makes the model reliable to also 
study the spectrum of the excited states. Indeed there is a clear way 
to associate the model states with the observed baryon excitations; 
except maybe an additional P11 nucleon state although there exist
analyses with such a resonance. Otherwise, this model calculation
did not indicate the existence of any so far unobserved baryon
state with quantum numbers of three quark composites.
The comparison with the empirical 
spectrum furthermore has led to the speculation that the so far 
undetermined spin--parity quantum numbers of the hyperon resonances
$\Xi(1690)$ and $\Xi(1950)$ should be $J^\pi=\frac{3}{2}^-$
and $J^\pi=\frac{1}{2}^+$, respectively. Lastly we have 
seen that the computed masses for the exotic $\Theta^+$
and $\Xi_{3/2}$ baryons nicely agree with the recent observation
for these pentaquarks. The quality of these results is remarkable, 
after all only at this stage the model contains no more adjustable
parameter. The mass difference between mainly octet
and mainly antidecuplet baryons thus is a prediction while it is an 
input quantity in most other approaches~\cite{Di97,Wa03,Bo03,Wu03,El04}.
We are thus confident that the present predictions for the masses of 
the spin--$\frac{3}{2}$ pentaquarks are sensible as well and we roughly 
expect them between $1.6$ and $2.1{\rm GeV}$. It is known and obvious
from the Hamiltonian that in chiral soliton model estimates the masses 
of states with otherwise identical quantum numbers decrease with spin. 
In the present calculation we have observed this pattern also for 
isospin.

Moreover, this approach almost naturally yielded a classification for 
the next to lowest--lying $J^\pi=\frac{1}{2}^+$ baryon states
that is similar to the $\mathbf{8}\oplus\overline{\mathbf{10}}$ 
scenario found in a diquark model for the pentaquark baryons. Even 
though our soliton model spectrum for these next to lowest--lying 
$J^\pi=\frac{1}{2}^+$ states qualitatively equals that of 
the diquark model, the structure of the eigenstates is quite
different; mainly because the wavefunctions contain significant
contributions from higher dimensional $SU_F(3)$ representations
as {\it e.g.} the $\mathbf{27}$--plet. We are thus led to the 
conclusion that a two component mixing scenario can only be
a first approximation to the description of pentaquark--like
baryon resonances.

The present treatment also allows one to address the large $N_C$ 
discussion that has been recently emerged for the study of pentaquarks 
in chiral soliton models. Vibrational excitation energies are 
${\cal O}(N_C^0)$, so are the $SU_F(3)$ rotational excitation energies. 
While the leading order cancels for mass differences between baryons 
built from $N_C$ quarks, it does not for exotics~\cite{Kl89} that 
have $N_C+1$ quarks and one antiquark.
This indicates that for the discussion of pentaquarks rotational and 
vibrational modes should be considered on equal footing and that 
mixing effects may become an issue. Since the transition 
operator between these modes originates from the flavor symmetry breaking 
part in the Lagrangian, the corresponding matrix element is potentially
small. Nevertheless the approximate equality of the excitation energies 
may trigger sizable mixing effects. In the vibrational treatment, the 
electric monopole (scaling) and the magnetic dipole (rotations) 
channels contribute to the scattering amplitude in the partial waves 
in which the low--lying $J^\pi=\frac{1}{2}^+$ resonances 
occur~\cite{Ha84b,Schw89}. In the present study we have considered
exactly these two degrees of freedom. However, we did not treat
them as RPA modes, {\it i.e.} ${\cal O}(\hbar\omega$) excitations, 
but went beyond such a quadratic approximation for the fluctuations
off the soliton. In this treatment we have indeed seen that the
leading excitation energies of these modes are almost identical. 
And, not surprisingly, the mixing is important for the 
excitations of octet and decuplet states. On the other hand the 
predicted masses for exotic states that do not have partners with 
equal quantum numbers in the octet or decuplet did not change 
significantly when we included the monopole degree of freedom. We 
therefore conclude that the purely rotational treatment of
such states reliably approximates their mass differences with
respect to the nucleon.

\acknowledgments
The author is grateful to G. Holzwarth, R. L. Jaffe, J. Schechter,
and H. Walliser for interesting discussions on the subject.

\end{document}